\documentclass[prd,tightenlines,showpacs,nofootinbib,amsfonts,amssymb,amsmath,color,11pt]{revtex4}
\usepackage{graphicx}
\usepackage{color}

\begin{document}
\newcommand{\etal}{{\it et al.}}
\newcommand{\bx}{{\bf x}}
\newcommand{\bn}{{\bf n}}
\newcommand{\bk}{{\bf k}}
\newcommand{\dd}{{\rm d}}
\newcommand{\dslash}{D\!\!\!\!/}
\def\ga{\mathrel{\raise.3ex\hbox{$>$\kern-.75em\lower1ex\hbox{$\sim$}}}}
\def\la{\mathrel{\raise.3ex\hbox{$<$\kern-.75em\lower1ex\hbox{$\sim$}}}}
\def\beq{\begin{equation}}
\def\eeq{\end{equation}}
\def\be{\begin{equation}} 
\def\ee{\end{equation}}
\def\bea{\begin{eqnarray}}
\def\eea{\end{eqnarray}}


\vskip-2cm
\title{Cosmic Acceleration from Abelian Symmetry Breaking}

\author{  
Gianmassimo Tasinato}
\affiliation{
 Institute of Cosmology \& Gravitation, University of Portsmouth, Dennis Sciama Building, Portsmouth, PO1 3FX, United Kingdom\\
}
\vspace*{2cm}

\begin{abstract}
 We discuss a consistent theory for a self-interacting vector field,  breaking  an Abelian 
symmetry in such  a way to obtain an interesting behavior  for its   longitudinal polarization. 
 In an appropriate decoupling limit,   the dynamics  of the longitudinal  mode is controlled by  Galileon interactions.
 The full theory away from  the      decoupling limit       does not propagate  ghost  modes, and     can be  investigated  in  regimes  
        where          non-linearities     become important.     When coupled to gravity, this theory  provides a candidate for dark energy, since 
 it admits  de Sitter   cosmological solutions  characterized by a technically natural value for the Hubble parameter.
     We also consider the  homogeneous  evolution   when, besides the vector,  additional matter in the form of perfect fluids is included.  We find       that the vector can have an important role in characterizing the    universe expansion.

\vspace{1cm}

\end{abstract}
 \date{February 2014}
 \maketitle

\section{Introduction}
 \label{sec:intro}
Imposing a gauge symmetry is a device  to remove degrees of freedom. The simplest example  
 is the  Abelian $U(1)$ gauge symmetry of electromagnetism.  Thanks to this symmetry, the vector associated with  a massless photon has two transverse polarizations only, while  its longitudinal polarization is absent from the spectrum of dynamical degrees of freedom.  Adding a mass
 term breaks the Abelian gauge symmetry, and makes  the longitudinal polarization a dynamical mode.
    In the limit of vanishing photon mass, the gauge symmetry is recovered 
   in its original form. Neglecting gravity, 
  the longitudinal polarization remains among the available degrees of freedom
  behaving as a free,  massless scalar field that does not interact with the transverse polarization modes.

    \smallskip
    
    But the degree of freedom eliminated by imposing an Abelian gauge invariance might not be so undesiderable after all. In some circumstances  
    it  can have interesting cosmological applications, as for example to provide 
    a natural candidate for      
      dark energy. 
    Is it possible   to break  the Abelian gauge symmetry acting on a vector field,
  in such  a way to
 get 
   a non-trivial  theory for its transverse polarization? 
 A    motivation for asking this question is an analogy with recent advances on massive gravity. In the dRGT model
  \cite{deRham:2010kj},  a  proper decoupling limit of vanishing graviton mass  leads to a rich  theory  for the graviton longitudinal  polarizations,  corresponding  to a combination 
 of Galileon Lagrangians \cite{Nicolis:2008in}. 
 Thanks to its connection with Galileons, 
   dRGT massive gravity is an appealing   set-up since it admits cosmological solutions describing accelerating universes
   in the vacuum \cite{deRham:2010tw,D'Amico:2011jj}, exhibits  a consistent realization of the Vainshtein screening mechanism \cite{Koyama:2011xz,Koyama:2011yg}, and keep
    quantum corrections under control in the regime of interest \cite{Nicolis:2004qq,deRham:2012ew}.
      See for example  \cite{oai:arXiv.org:1105.3735,Tasinato:2013rza,deRham:2014zqa} for recent reviews on massive gravity.

\smallskip

As we will discuss in this paper, an analogous  situation  can be obtained in a simpler theory of self-interacting spin one vector fields,
 with broken Abelian gauge symmetry. When suitable derivative self-interactions are included, 
    the dynamics of the vector longitudinal degree of freedom  is  non-trivial. In an appropriate decoupling limit, the theory 
    recovers the Abelian gauge invariance, and the dynamics of the 
     vector longitudinal  polarization is controlled by a combination of Galileon  Lagrangians. The full theory away from  the 
     decoupling limit is consistent, in the sense that
      it does not propagate an additional ghostly fourth mode. 
       The system can be  investigated  in non-trivial regimes  
        where the effects of non-linear interactions    become important.
       When coupled to gravity, it admits cosmological solutions 
   describing  accelerating universes with no need of an additional energy momentum tensor,  providing
   a candidate  for dark energy     
   with a technically natural size for the dark energy scale. 
      Moreover,  when 
   adding on top of the vector content a combination of  perfect fluids 
   with constant equation of state,
    the resulting cosmological expansion is characterized by 
    a Friedmann equation  with peculiar properties and
    with potentially interesting cosmological consequences.  Indeed,
     we find  that the vector can have an important role in characterizing   gravitational interactions around 
     cosmological backgrounds, and the cosmological expansion of our universe.

  \smallskip
  
     The vector with broken gauge symmetry we are considering is not necessarily the photon.
      For simplicity, we can regard it as an additional field with no  direct couplings to Standard Model particles, although
      as we will briefly discuss 
       the parameters in our scenario might be accommodated to satisfy the existing bounds. 
         In the past, many scenarios have been considered  
        for  modifying General Relativity through  
        the dynamics of vectors,
         with important  cosmological consequences for dark energy and dark matter. The first working models were  introduced in
        the early seventies 
        by Will, Nordtvedt, Hellings \cite{Will:1972zz,Hellings:1973zz}; more recently, 
         well studied proposals have been the Einstein-Aether theory \cite{Jacobson:2000xp}
          and the TeVeS covariantized version of MOND \cite{Bekenstein:2004ne}. 
           See \cite{Clifton:2011jh} 
       for a comprehensive review  with a complete list of references to the relevant literature. 
         The novelty of our approach is the 
         emphasis  on symmetry arguments for building  our theory,
          so to obtain
                              a compact structure for our Lagrangian   that  
          makes
          explicit 
          connection with Galileons. This fact can allow us 
           to keep
         our set-up 
          under control  in   strong coupling  
           regimes, where potentially interesting effects occur.

  \smallskip

      The paper is organized as follows.  We start in Section \ref{sec-setup}  describing 
      how our vector Lagrangian is built, and discussing  its physical features including the connection
      with Galileons. We continue with 
         Section \ref{sec-cosmo}, investigating
 applications to cosmology. We conclude  in Section \ref{sec-disc}.
   
     \section{The set-up}\label{sec-setup}
     
  Consider the following  vector Lagrangian in Minkowski space (adopting the mostly plus signature), for the moment ignoring any coupling with gravity
\be\label{vtlag}
{\cal L}\,=\,-\frac14\,F_{\mu\nu} F^{\mu\nu}+
\sum_{i=0}^3\,
{\cal L}_{(i)}\,,
 \ee
with $F_{\mu\nu}\,=\,\partial_\mu A_\nu-\partial_\nu A_\mu$, and $A_\mu$ a  vector field.  
The  symmetry-breaking Lagrangians  ${\cal L}_{(i)}$ 
we consider, besides the usual Proca mass term, 
are defined in terms of derivative self-interactions of the vector as
\bea
{\cal L}_{(0)}&=&-{m^2}\,A_\mu A^\mu\,,\\
{\cal L}_{(1)}&=&-{\beta_2}\,A_\mu A^\mu\,
\left( \partial_\rho A^{\rho}\right)
  \,,
\\
{\cal L}_{(2)}&=&-\frac{\beta_3}{m^2}\,A_\mu A^\mu\,\left[
\left( \partial_\rho A^{\rho}\right) \left( \partial_\nu A^{\nu}\right) 
-\left( \partial_\rho A^{\nu}\right) \left( \partial^\rho A_{\nu}\right) 
 \right]\,,
\\
{\cal L}_{(3)}&=&-\frac{\beta_4}{m^{4}}\,A_\mu A^\mu\,
\Big[
-2\left( \partial_\mu\,A^\mu\right)^3+3\,\left( \partial_\mu\,A^\mu\right)\,\left( \partial_\rho\,A^\sigma \partial^\rho A_\sigma\right)
+3 \,\left( \partial_\mu\,A^\mu\right)\,\left( \partial_\rho\,A^\sigma \partial_\sigma A^\rho\right)
\nonumber\\
&&-\,  \partial_\mu\,A^\nu\,\partial_\nu\,A^\rho\,\partial_\rho\,A^\mu-3\,\partial_\mu\,A^\nu\,\partial_\nu\,A^\rho\,\partial^\mu\,A_\rho
\Big]\,,
\eea
 and  break the Abelian gauge symmetry
$A_\mu\,\to\, A_\mu+\partial_\mu \xi$.  Here, $m$ has dimension of a mass, while the $\beta_i$ are dimensionless couplings. 
The suffix $(i)$ in the Lagrangians indicates the number of derivatives in each term.  Notice that these interactions   do
not break  
 Lorentz symmetry, in particular 
   they 
   do not select any preferred frame.  
The Lagrangians ${\cal L}_{(i)}$ are built by the following combinations made with antisymmetric $\epsilon$ tensors
 in four dimensions
\be\label{genelag}
{\cal L}_{(i)}\,\propto\,A_\mu A^{\mu}\,\left(\epsilon_{\alpha_1\,\dots\,\alpha_{i} \gamma_{i+1}\,\dots\,\gamma_4} 
\epsilon^{\beta_1\,\dots\,\beta_{i} \gamma_{i+1}\,\dots\,\gamma_4} \, \partial_{\beta_1 }A^{\alpha_1}\dots  \partial_{\beta_i }A^{\alpha_i}
\right)\,.
\ee
 These derivatives self-interactions are chosen in such a way as to lead to a consistent set-up, in the sense
that a fourth `ghost-mode' cannot be excited.    Indeed,
 it is simple to show that, due to the antisymmetric properties
of the $\epsilon$ tensor, the Lagrangians ${\cal L}_{(i)}$ {\it do not} contain contributions containing time derivatives of  the time component $A_0$ of the vector
(up to total derivatives): hence the equation of motion for this component is a constraint equation. On the other hand, the
Lagrangians ${\cal L}_i$ 
 break the  Abelian
  gauge symmetry:  the theory contains {\it three} dynamical modes, the usual transverse plus the longitudinal polarization of 
 the vector.
 As we will see,
   the latter degree of freedom, when $m^2>0$, is well behaved. So, we end up with a consistent theory with three healthy modes
  around Minkowski space. 
  
 \bigskip
 
 In what follows, we would like to investigate the interesting dynamics of the
  vector longitudinal polarization associated with the previous Lagrangians.

\subsection{Vector field produced by a static source}

 For simplicity,  
  in this subsection 
 we  
  include (besides the standard kinetic term)   the Lagrangians ${\cal L}_{(0),\,(1)}$ only.
 Hence the Lagrangian on which we now focus our attention  is 
  \be
  {\cal L }_T\,=\,-\frac14\,F_{\mu\nu} F^{\mu\nu}-m^2 A_\mu A^\mu-
{\beta}\,A_\mu A^\mu\,
\left( \partial_\rho A^{\rho}\right)\,.
\label{attlag}
\ee
 To gain some initial flavor of the physical  effects associated with the non-linear  self-couplings of the vector, let us   
 analyze a static system
 of a   charged density with associated current $J^{\mu}\,=\,\left( \rho, \,0,\,0,\,0\right)$, minimally 
 coupled to the vector with a term $A_\mu\,J^\mu$ in flat space.  
    We would like to 
  write
   the equations corresponding to 
    a vector 
    field configuration produced by such a  body. 
  We focus on static configurations: $A_\mu\,=\,A_\mu(0,\vec{x})$, and split the vector potential in components as $A_\mu\,=\,\left( A_0,\,A_i\right)$. 
  The equations of motion for the vector degrees of freedom are
\bea
-\vec{\nabla}^2\,A_0
&=&\rho-2\,m^2\,A_0-2\,\beta\,A_0\,\partial_i A_i
 \,,\\
2 m^2 A_i&=& 
\vec{\nabla}^2\,A_i-\partial_i \partial_j A^j+ \beta\,\partial_i \left( -A_0^2+A_j^2\right)
-2 \beta A_i\,\partial_j A_j\label{eqm2}\,,
\eea
with $\vec{\nabla}^2\equiv \delta_{ij} \partial_i \partial_j$. The main difference
with respect to the  gauge invariant (and Proca) cases is  that the $\beta$ contribution renders the $A_0$ equation dependent
on the quantity $\partial_i A_i$. Taking the divergence  of eq (\ref{eqm2}), we find
\be
2\,m^2\,\partial_i A_i\,=\,-{\beta}\,\nabla^2\,A_0^2
-2\beta\left( \partial_i A_i\,\partial_j A_j-\partial_i A_j\,\partial_j A_i\right)\,.
\ee
 In looking for a static field configuration,   we separate the
 spatial vector components in transverse and longitudinal parts, $A_i\,=\,A^T_i+\partial_i \,\chi$ with $\partial_i A_i^T\,=\,0$. 
 We focus  here on a simplifying 
   Ansatz 
     setting  to zero the transverse polarizations $A^T_i=0$. Hence 
  we end up with the coupled equations for $A_0$ and $\chi$
  \bea
-\vec{\nabla}^2\,A_0
&=&\rho-2\,m^2\,A_0-2\,\beta\,A_0\,\vec{\nabla}^2\chi\,,\label{eqp1}
 \\ \label{eqp2}
\vec{\nabla}^2\chi&=&-\frac{\beta}{2\,m^2}\,\vec{\nabla}^2\,A_0^2-\frac{\beta}{2\,m^2}\,\left[ \left(\vec{\nabla}^2\,\chi \right)^2
-\left( \partial_i \partial_j\,\chi\right)^2
\right]\,.
 \eea
 Notice that, although the longitudinal polarization $\chi$ is not directly coupled to the source, nevertheless it `feels' it via 
 the non-linear term in eq. (\ref{eqp1}). 
 Let us make the further simplifying Ansatz of spherical symmetry, 
       where all the functions
      depend only on the distance  $r$ from the origin, and the previous two equations (\ref{eqp1}-\ref{eqp2}), 
      after some manipulations,
      read
      \bea
-\frac{d}{d r}\left( r^2 A_0'\right)&=&r^2\,\rho-2\, m^2\,r^2\, A_0+2\beta\,A_0\,
\frac{d}{d r}\left( r^2 \chi'\right) \label{eqp1a}\,,\\
 \chi'&=& \frac{2\,\beta}{m^2}\,\frac{\chi'^2}{r}+\frac{\beta\,  A_0 A_0'}{ m^2}\,,
\label{eqp2a}
      \eea 
      where a prime indicates derivative along $r$.
      Eq. (\ref{eqp2a}) is a second
      order algebraic equation for $\chi'$, whose solution  provides a relation between $\chi$ and $A_0$ (we focus
      only on the branch  that decays for large values of $r$):
      \be\label{solchi}
      \chi'\,=\,\frac{m^2\,r}{4\beta}\,\left(1-\sqrt{1-\frac{8 \,\beta^2\,A_0 A_0'}{m^4\,r}} \right)
     \,. \ee
      This relation can be substituted in eq (\ref{eqp1a}) to obtain a non-linear differential  equation 
       that govern the behavior of the `electric field' produced by the source.
           At large distances from the source,
           where  $A_0$ is small, eq. (\ref{solchi}) can be expressed as 
           $$
             \chi'\,\simeq\,\frac{\beta}{m^2}\,A_0 A_0'
           $$
          and one finds that   both  $A_0$ and $\chi$ acquire a Yukawa-like suppression (we normalize to unity the  charge of the source):
          \bea
          A_0&\simeq&\frac{e^{-\sqrt{2}\,m\,r}}{r}\,,\label{ldA0}\\
          \chi&\simeq&\frac{\beta\, e^{-2\sqrt{2}\,m\,r}}{2\,m^2\,r^2}.\label{ldchi}
          \eea
           Notice  that $\chi$ decays more rapidly than $A_0$. We call $r_m\,\equiv\,1/\left( \sqrt{2}\, m\right)$
           the distance at which the Yukawa-like behavior due to the vector mass  becomes important in determining  the profile for $A_0$:
               well
                below this radius, the solution for the vector potential, eq  (\ref{ldA0}),  can be approximated by a power-law. 
            In this regime $r\ll r_m$, one can identify another characteristic distance, corresponding to the `strong coupling' scale at which 
            the argument in the square root in eq. (\ref{solchi}) becomes appreciably different than one: this scale is given by 
            \be
            r_s\,\equiv\,\frac{\sqrt{\beta}}{m}\,.
            \ee
            By choosing $\beta$ sufficiently small, $r_s$ 
             can be made parametrically smaller than 
            $r_m$.   The regime $r_s\ll r \ll r_m$ is interesting since the non-linear 
             contributions weighted by $\beta$ in eq (\ref{eqp1a}) can be neglected, as well as the mass term, and the power-law configurations
              $A_0\sim 1/r$, $\chi\sim r_s^2/r^2$ are
              solutions for  the equations of motion.  It is  an intermediate  regime 
                in which, although $\chi$ acquires
 a non-trivial profile
  due to the non-linear interactions weighted by $\beta$, its effect is too weak to appreciably influence  the configuration for $A_0$.  It would be interesting to numerically investigate
 the full strong coupling regime $r\ll r_s$,
 in particular  to understand whether interesting
   screening effects on this vector set-up
   appear,
 similarly 
 to what happens for the  gravitational Vainsthein effect \cite{Vainshtein:1972sx}.

\subsection{Relation with scalar Galileons}\label{relgal}

       That some interesting non-linear regime exists nearby a source is suggested by observing that   
the non-linear       equations  (\ref{eqp1}-\ref{eqp2})  preserve a (spatial) Galileon symmetry in the 
 longitudinal polarization, $\chi\,\to\,\chi+a+b_i x_i$, and  Galileon systems are known to exhibit 
 a screening 
 Vainshtein mechanism \cite{Nicolis:2008in} in  gravitational set-ups. Indeed,
 our motivation for presenting the non-linear coupled equations sourced by a static charge  was precisely to 
 point out this fact. We now investigate  
   in more detail  how the vector Lagrangian (\ref{attlag})  is  connected with Galileons. 
  We adopt the St\"uckelberg formalism,
   trading everywhere
  $A_\mu$ for $A_\mu+1/\left(\sqrt{2} \,m\right) \,\partial_\mu \phi$:  the resulting Lagrangian
 is invariant under the  gauge symmetry $A_{\mu}\,\to\,A_\mu-\partial_\mu \xi$, $\phi\to\phi+\sqrt{2}\,m\,\xi$. The
 scalar field $\phi$ 
  plays the same physical role as that of the longitudinal vector polarization.   The use of the St\"uckelberg approach
  renders clearer  the interactions among the different degrees of freedom. 
  The total Lagrangian reads, assuming  $m^2>0$ to avoid ghost instabilities,
   \bea
{\cal L}_{T}&=&-\frac14\,F_{\mu\nu} F^{\mu\nu}-\,\frac12\,\left(\sqrt{2}\,m\,A_\mu+\partial_\mu \phi\right) \left( \sqrt{2}\,m\,A^\mu+\partial^\mu \phi \right)\nonumber\\
 &-&\frac{{\beta}}{\sqrt{8}\,m^3}\,\left(\sqrt{2}\,m\,A_\mu+ \partial_\mu \phi\right)  \left( \sqrt{2}\,m\, A^\mu+
\,
\partial^\mu \phi \right)\,\left(\sqrt{2}\,m\, \partial_\nu A^\nu+\,\partial_\nu \partial^\nu \phi\right)\,.\eea
 To isolate the (self-)interactions of the St\"uckelberg field $\phi$ we take the `decoupling' limit
\be
m\to 0 , \,\,\beta\,\to\, 0, \,\hskip1cm\frac{\beta}{m^3}\,=\,{\rm fixed}\,=\,\frac{\sqrt{2}}{\Lambda_G^{3}}\,, 
\label{cutoffG}\ee
leading to 
 \bea
{\cal L}_{dec}&=&-\frac14\,F_{\mu\nu} F^{\mu\nu}-\,\frac12\,\partial_\mu\phi \partial^\mu \phi
 -\frac{{1}}{2\,\Lambda_G^3}\,\left(\partial_\mu \phi   
\,
\partial^\mu \phi \right)\,\partial_\nu \partial^\nu \phi\,.\eea

The result of taking  such a decoupling limit is a
 theory with {\it two} different symmetries~\footnote{Analogous
   arguments straightforwardly apply also to the complete set of interactions ${\cal L}_{(i)}$ in eq. (\ref{vtlag}),
   leading to higher order scalar Galileon Lagrangians.}: 
a  free vector Lagrangian that satisfies the Abelian gauge symmetry, plus a cubic Galileon   scalar Lagrangian 
  controlled by the 
strong coupling 
scale $\Lambda_G$, and
respecting a Galileon symmetry $\pi\to\pi+b+a_\mu x^\mu$.  
 This feature makes stable the size
 of the parameters $m$ and $\beta$, 
   since  keeping them small is technically natural in the 't Hooft sense \cite{'tHooft:1979bh}.
  It would  also be interesting 
 to analyze in detail  the issue
 of quantum corrections to this set-up. In particular, to try    to understand whether
  additional operators --   that would  spoil the structure of our Lagrangian -- can be kept under control 
  when 
  working
   in some strong or intermediate coupling  regimes, in analogy  
  with what   happens for Galileons or massive gravity \cite{Nicolis:2004qq,deRham:2012ew}.
   Related to this, it would be interesting to understand whether conformal versions of this vector Lagrangian
   can be constructed, using for example the methods of \cite{Tasinato:2013wna}, to find relations with conformal 
   Galileon theories \cite{Nicolis:2008in}. 
 
Moreover, 
the connection we found with Galileons 
 provides another perspective on    why
  the theory under consideration is consistent (ghost free) around Minkowski space, and
  promises to lead to  interesting cosmological applications as accelerating configurations.  

\subsection{Coupling to gravity}

Coupling  our theory  to gravity presents the very same issues one meets in the covariantization of scalar Galileon theories.  
In order not to propagate ghosts, we require that our set-up
  does not lead to   derivatives higher  than two in the equations of motion for vector and gravitational degrees
of freedom. Applying for example  the approach
   developed in \cite{Deffayet:2009wt,Deffayet:2009mn},  one  finds  a consistent covariantization of the
Lagrangian densities ${\cal L}_{(1)}$, ${\cal L}_{(2)}$: 
\bea
{\cal L}^{cov}_{(1)}&=&-{\beta_1}\,A_\mu A^\mu\,
\left( \nabla_\rho A^{\rho}\right)
  \,,
\\
{\cal L}_{(2)}^{cov}&=&-\frac{\beta_2}{m^2}\,A_\mu A^\mu\,\left[
\left( \nabla_\rho A^{\rho}\right) \left( \nabla_\nu A^{\nu}\right) 
-\left( \nabla_\rho A^{\nu}\right) \left( \nabla^\rho A_{\nu}\right) 
 -\frac14\,R\,A_\sigma A^\sigma\right]\,,\label{cova2}
\eea 
with $\nabla_\mu$ the usual covariant derivative in curved space, and $R$ is the Ricci scalar. 
 Notice that the vectors couple non-minimally to gravity, thanks to the coupling with the Ricci scalar in eq. (\ref{cova2}). 
 For our purposes, we will not need to covariantize ${\cal L}_{(3)}$: this is left for future work. It is simple to check
   that in an appropriate
 decoupling limit (as discussed in subsection \ref{relgal}) the previous formulae reduce to the covariantized cubic and quartic
 scalar Galileon Lagrangians. 
 It would be interesting to analyze whether the vector interactions 
       can contribute to a gravitational Vainshtein mechanism around a spherically  symmetric source, 
       as investigated for   a scalar-vector set-up in \cite{Tasinato:2013oja}.      
 
 \smallskip
 Armed with these results,
  we will  now 
  focus on the action
 \be
 {\cal S}\,=\,\int d^4x\,\sqrt{-g}\,\left[\frac{M_{Pl}^2}{2}\,R-\frac14 F_{\mu\nu} F^{\mu\nu}-m^2\,A_\mu A^\mu+{\cal L}^{cov}_{(1)}+{\cal L}^{cov}_{(2)} \right]
 \ee 
  with the aim to study its cosmological implications.


\section{Applications to cosmology}\label{sec-cosmo}

 We consider  a homogeneous  
  FRW metric
 with flat spatial curvature 
\be
d s^2\,=\, -d t^2 +a^2(t)\, 
\delta^{ij}\,d x_i \,d x_j
\ee
with $a$ the scale factor, and $H=\dot{a}/a$ the corresponding Hubble parameter. 
The vector potential is $A_\mu\,=\,\left(A_0,\,A_i\right)$. The spatial vector components are
  decomposed in $A_i\,=\,A_i^T+\partial_i \,\chi$
with
$\partial_i A_i^{T}\,=\,0$. 
  We investigate  homogeneous configurations. 
 We consider a background vector profile with only the time-component
turned on:
 $A_\mu\,=\,\left( A_0 (t)\,,0\,,0\,,0\right)$. We avoid to turn on spatial components for the vector to avoid anisotropies and the 
corresponding  generic  instabilities pointed out in \cite{Himmetoglu:2008zp}. 
 The   equation of motion for $A_0$ is a constraint equation, since the Lagrangian does not depend
  on time derivatives of $A_0$,  and reads
$$
A_0\left(m^2  -3\,\beta_1\,A_0\,H+9 \,\frac{\beta_2}{m^2}\,A_0^2\,H^2\right)\,=\,0
\,.$$

 We can identify various
   branches of solutions: one is the trivial $A_0\,=\,0$, while the most interesting ones for us are
 \bea 
 A^{\pm}_0(t)&=& \frac{\beta_1\pm\sqrt{\beta_1^2-4 \beta_2}}{6\,\beta_2} \,\frac{m^2}{H(t)}\,,\label{solA0}
 \\
 &=&\frac{c_{\pm} \,m^2}{H(t)}\,.\label{solA0b}
 \eea
 These branches  require $\beta_1^2\ge4\beta_2$ to have a  real square root.  
 In the second line we defined the dimensionless parameters $c_\pm$ built in terms of $\beta_1$, $\beta_2$.
 From now on, for definiteness, we will focus on the case $\beta_1\ge0$, $\beta_2\ge0$. 
  Using the non-trivial solutions  (\ref{solA0b}) for $A_0$, one finds that  
 the  content of the energy momentum tensor has a perfect fluid structure, 
   with vector energy density and pressure given by
 \bea
 \rho_{\small{V}}&=&\frac{c_\pm^2\left( 9\beta_2 c_\pm^2-2\right)\,m^6}{2H^2}\,,
 \\
 p_V&=&\frac{c_\pm^2\left( 2-9\beta_2 c_\pm^2\right)\,m^6}{2H^2}+\frac{c_\pm^3\,\left(
 9 \beta_2 c_\pm-2 \beta_1
  \right)\,\dot{H}}{H^4}\,.
 \eea
  It is simple to show that, in order to have a positive vector energy density, $\rho_V\ge0$, one has to focus
  on the positive  branch of solutions in eq (\ref{solA0}), that require a non-vanishing $\beta_2$.   
  The Friedmann equation reads
  \be
  H^2\,=\,\frac{c_{\pm}^2 \left( 9\, \beta_2\,  c_{\pm}^2-2\right)\,m^6}{6\,H^2\,M_{Pl}^2}\,,
  \ee
  that is solved for a constant Hubble parameter. A real solution for the scale factor can be found focussing on the 
   positive  branch of  eq (\ref{solA0}),  where the (square of the) Hubble parameter results 
 \bea\label{valH}
 H^2&=& \left(\frac{c_+}{\sqrt{6}}\,\sqrt{9\, \beta_2\,  c_{+}^2-2} \right)\,\frac{m^3}{M_{Pl}}
 \eea
 and is well defined when 
  $\beta_1^2\,>\,9\,\beta_2/2$, a condition that we 
will impose from now on. The overall
 dimensionless  coefficient in front of the right hand side of the previous equation -- call it $c_\beta$ --
 simplifies in the small $\beta_2$ limit, reducing to
$c_\beta\simeq\beta_1^2/\left( 108\,\beta_2^{3}\right)^{1/2}$.

\smallskip

  
Hence, the dynamics associated with  the new vector interactions
 is able to  
drive cosmological acceleration with a constant (de Sitter) equation of state.  At the background level, 
 such 
cosmological 
acceleration  is
identical to the one driven by
a positive
  `cosmological constant' of size 
\be\Lambda_V^4\,=\,6\,{c_\beta}\,{m^3\,M_{Pl}}\, \label{efflv}\ee 
 where the quantity $\Lambda_V$ has the  dimension of a mass, 
 and allows us 
    to  write more concisely  $H^2\,=\,\Lambda^4_V/(6\,M_{Pl}^2)$. 
  In order to be able to drive a de Sitter expansion with  the current value for the Hubble parameter,
 the mass parameter $m$  should be chosen to be of  order
\be
m\,\simeq\,c_\beta^{-1/3}\,10^{-13}\,eV\,.
\ee
The current limit on the photon mass is $m_{\gamma}\le 10^{-18} \,eV$ \cite{Beringer:1900zz}, that could be satisfied by taking a sufficiently
small value for $c_\beta$.  
  Small values for our parameters are technically natural in the 't Hooft sense, since as sending $m$ (and the $\beta_i$) to zero one 
   recovers  Abelian and Galileon symmetries. Hence, although we are keeping
   our discussion completely general,  one
    might think to use the photon itself as the self-interacting vector we are investigating.  Let us point out that
    the non-linear vector interactions we are analyzing, with their associated  strong coupling effects, can considerably affect the existing bounds:
    see \cite{Goldhaber:2008xy}  for a critical discussion on photon mass limits.
   
  \smallskip
  
  It would also be interesting to study in detail the dynamics of cosmological perturbations around
   the time dependent configurations we have presented.
    We leave 
 this task for future work, but let us mention that we checked that,
   after 
   including  the contributions
  from  the homogeneous background,  the effective mass 
  parameter for the transverse  vector fluctuations   $A_i^T$  does 
   maintain the correct sign around  this cosmological solution.
    
  \smallskip

\bigskip
 Let us investigate  a bit further   the background homogeneous cosmology in our set-up.  We will
 see that vectors can have an interesting role in characterizing the cosmological evolution. 
 On top of the vector content
  previously analyzed, we include additional matter content  in the form
 of     perfect fluids with constant equation of state,  with total energy density $\rho$,
  for simplicity not directly coupled to the vector. The 
   first Friedmann equation  now reads
\be
 H^2(\tau)\,=\,\frac{\rho}{3\,M_{Pl}^2}+\frac{\Lambda_V^8}{36\,H^2(\tau)\,M_{Pl}^4}\label{Hequa}
 \ee
 with $\Lambda_V^4$ the effective cosmological constant induced by the vector, as defined in equation (\ref{efflv}). The second contribution is peculiar, 
 since it contains an $H^2$ in the denominator.
  Eq. (\ref{Hequa}) can be solved expressing the Hubble parameter in terms of
   the remaining quantities: the branch of solutions corresponding to a real $H$ is
   
   \be
H^2\,=\,   \frac{\rho+\sqrt{\rho^2+\Lambda_V^8 }}{6\,M_{Pl}^2}\,.
   \label{solh2}
   \ee 
   Such `Friedmann-like' equation has a non-standard structure, due to the square root in the right hand side.  
   Interestingly, it  admits solutions also for a negative energy density $\rho$ (for
   example, a $\rho$ dominated by a negative 
  bare cosmological constant) in absence of spatial curvature. 
  
  The standard form of the Friedmann equation -- in absence of a cosmological constant -- is obtained in the limit $\rho \gg
   \Lambda_V^4$.  In the opposite limit, $\rho\ll \Lambda_V^4$, we 
   expand (\ref{solh2})
    obtaining
   \be
   H^2\,=\,\frac{\rho}{6 M_{Pl}^2}+\frac{\Lambda_V^4}{6\,M_{Pl^2}}+\frac{\rho^2}{12\,\Lambda_V^4\,
   M_{Pl}^2}+\dots
   \ee
The linear term in $\rho$ differs from  the standard form for the Friedmann equation due to the factor of two in the
denominator. This suggests that, in this small $\rho$ limit,
  the effective Newton constant in this cosmological background is half  the one in Minkowski space, in other words 
  $M_{Pl}^{cosm}\,=\,\sqrt{2}\,M_{Pl}^{Mink} $. (We checked that the same behavior occurs for the second
  Friedmann equation,   governing the
  second time derivative of the scale factor.)  Hence, vector degrees of freedom play a relevant role in characterizing gravitational interactions and cosmological evolution 
  around non-trivial backgrounds, since they `renormalize' the value of the Newton constant.
  This fact could also be argued    from the structure of 
   our covariantized action in
   equation (\ref{cova2}), where we learn that vectors are non-minimally
     coupled with  the Ricci scalar. This
      implies that the dimensionful  coefficient in front of the Ricci scalar  in the action -- that sets the strength
      of gravitational interactions -- can depend on the vector background. 
       It would be very interesting to directly calculate the gravitational force between test bodies
   in these cosmological configurations, to  understand more explicitly  
   the role of vectors in determining the   gravitational  
   force. 
  
  \section{Discussion}\label{sec-disc}

In this work we discussed a consistent  theory for a self-interacting vector field  that breaks an Abelian 
symmetry, in such  a way to obtain an interesting dynamics for the vector  longitudinal polarization. In an appropriate
decoupling limit, 
 the dynamics  of the longitudinal scalar mode is controlled by  Galileon Lagrangians.
 The full theory away of the 
     decoupling limit is consistent in the sense that
      it does not propagate a  ghostly fourth mode. 
        The system can be  investigated  in non-trivial regimes  
        where the effects of non-linear interactions    become important.
       When coupled with gravity
 it admits a de Sitter branch of  cosmological solutions  characterized by a technically natural value for the Hubble parameter.
    We  studied the homogeneous cosmological evolution
  when additional matter in the form of perfect fluids is included in the energy momentum tensor. 
   The resulting cosmological expansion is characterized by 
    a Friedmann equation  with peculiar properties and
    with potentially interesting cosmological consequences.  Indeed,
     we found  that the vector can have an important role in characterizing   gravitational interactions around 
     cosmological backgrounds, and the cosmological expansion of our universe.


\smallskip

As mentioned above,
 the non-linear self-interactions for the transverse vector polarizations are controlled by  Galileon combinations; hence,
 strong coupling effects can play a role in physically interesting situations.  
 The relation 
  with Galileon and Abelian symmetries in appropriate limits  renders the theory technically natural, 
  allowing to keep 
   the size of the available parameters under control.
      It would be interesting  to further explore   our theory in non-linear regimes to understand whether the particular
   structure we have chosen for our Lagrangian remains valid when quantum corrections are taken into account.
    Also, on a more phenomenological side,  it will be important to investigate in more details the accelerating  cosmological
    configurations we have determined, in particular the stability of fluctuations around them.  
   
   \smallskip
   
      While in this work we  did not specify the microscopic  nature of the  vector,  it will be interesting to explore
  more in detail whether the photon can play its role.
   We have explained that current photon mass limits can be satisfied by a suitable and technically natural choice of the 
   available parameters. On the other hand,  it is very likely that  the
    non-linear interactions we have analyzed considerably
        affect
   the existing bounds.
         Besides cosmology,
   it would  also be interesting  to investigate whether our  
   interactions can be
     obtained via a Higgs mechanism, 
and whether they can be      
    realized in some specific condensed matter physics set-up where Abelian symmetries
   are spontaneously broken.

\acknowledgments
It is a pleasure to thank Marco Crisostomi, Matthew Hull,  Kazuya Koyama,  Gustavo Niz, Ivonne Zavala for useful comments on the draft, and STFC for financial support through 
the grant  ST/H005498/1.

\end{document}